\documentclass[%
 aip,
 jmp,%
 amsmath,amssymb,
 reprint,%
]{revtex4-1}

\usepackage{graphicx}
\usepackage{dcolumn}
\usepackage{bm}

\begin{document}


\title{Photoacoustic ultrasound sources from diffusion-limited aggregates}

\author{Krutik Patel}
\affiliation{Department of Physics, University of California, Santa Barbara, California 93106, USA}
\author{Morgan Brubaker}
\affiliation{Department of Physics, University of California, Santa Barbara, California 93106, USA}
\author{Alexander Kotlerman}
\affiliation{Department of Physics, University of California, Santa Barbara, California 93106, USA}
\author{Robert Salazar}
\affiliation{Department of Physics, University of California, Santa Barbara, California 93106, USA}
\author{Eli Wolf}
\affiliation{Department of Physics, University of California, Santa Barbara, California 93106, USA}
\author{David M. Weld}
\thanks{Corresponding author}
\email{weld@physics.ucsb.edu}
\affiliation{Department of Physics, University of California, Santa Barbara, California 93106, USA}

\date{\today}

\begin{abstract}
Metallic diffusion-limited aggregate (DLA) films are well-known to exhibit near-perfect broadband optical absorption. We demonstrate that such films also manifest a substantial and relatively material-independent photoacoustic response, as a consequence of their random nanostructure.  We  theoretically and experimentally analyze photoacoustic phenomena in DLA films, and show that they can be used to create  broadband air-coupled acoustic sources.  These sources are inexpensive and simple to fabricate, and work into the ultrasonic regime.  We illustrate the device possibilities by building and testing an optically-addressed acoustic phased array capable of producing virtually arbitrary acoustic intensity patterns in air.
\end{abstract}

\maketitle

\section{Introduction}
Stochastically nanostructured materials exhibit optical, electronic, and thermal properties which can differ drastically from those of ordered nanostructures and of homogeneous samples of the same material.  In nanostructured aggregates, a combination of strong broadband optical absorption and fast local thermal response gives rise to a strong photoacoustic effect, in which amplitude-modulated light is transduced to sound at the frequency of amplitude modulation.  In this work we theoretically and experimentally investigate the photoacoustic response of nanostructured aggregates in air, and demonstrate that such materials enable the construction of  acoustic sources including an optically-addressed photoacoustic phased array.  The flexibility of such acoustic sources could enable low-cost generation of arbitrary broadband acoustic and ultrasonic waveforms, with potential applications ranging from subdiffraction air-coupled acoustic sources to materials characterization.

The study of  photoacoustic phenomena in bulk materials has an illustrious history~\cite{bell-photoacoustic}, and the effect has proven useful as a characterization tool in a variety of contexts~\cite{VARGAS198843}.  Photoacoustic material response has also been explored as an optically-controlled source of sound waves: for example, photoacoustic sources have been fabricated from lithographically-defined nanostructures, thin films, multilayer polymer-metal composites, nanoparticle-filled epoxies, and carbon nanotubes~\cite{Baac2012,Houe2008,Biagi2001,odonnell-thinfilmPA,odonnell-nanoPA}. In this work, we discuss a simple and low-cost technology for air-coupled photoacoustic ultrasound generation based on DLA films.  The technology is within the capabilities of nearly any laboratory, requiring only very moderate (few torr) vacuum conditions, inexpensive metal sources, and simple modulated LEDs.  We illustrate  potential applications by constructing an optically-addressed photoacoustic phased array capable of producing patterned ultrasonic  fields at far lower complexity and cost than a conventional multiple-piezoelectric-source phased array.  

\section{Sample Fabrication \& Morphology}
Highly disordered nanostructured films of many evaporable substances can be simply and reproducibly prepared by thermal evaporation in the presence of a low-pressure background of inert gas.  In films prepared using this procedure, a complex granular structure results from the randomization of atomic velocities before aggregation in the gas phase~\cite{pfund-metalblacks,broida-aggregates, goldblack1948, goldblack1993, goldblack2003}.   Such films often but not always resemble the product of diffusion-limited aggregation.  Diffusion-limited aggregation occurs when  particles executing a random walk collide and stick together; this process occurs in a wide variety of physical contexts, and is expected to produce a treelike fractal structure~\cite{DLA-original}.  As a concrete example, we find that nanostructured bismuth films are reliably produced in a standard thermal evaporator by running 115 A of current through a 9 m$\Omega$ alumina-coated molybdenum evaporation boat for 4 minutes, in a background pressure of 2 Torr of Argon.  The nanostructured films are deposited on any surface above the boat, although for surfaces too close to the boat they tend to melt and coarsen.   The  properties of the film do not  depend sensitively on deposition parameters.  Material costs for such a deposition are on the order of a few cents for a common evaporant like bismuth, and the total time to produce a film is less than ten minutes once the system is pumped down.
\begin{figure}[t]
\begin{center}
\includegraphics[width=0.48\columnwidth]{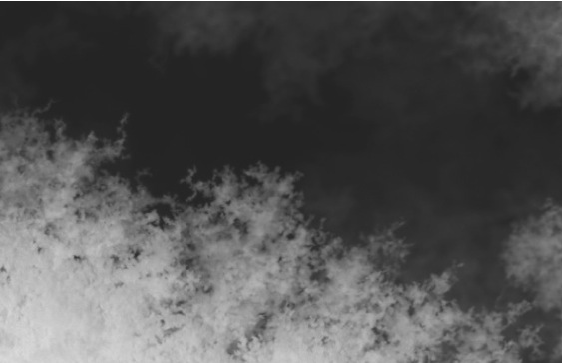}
\includegraphics[width=0.48\columnwidth]{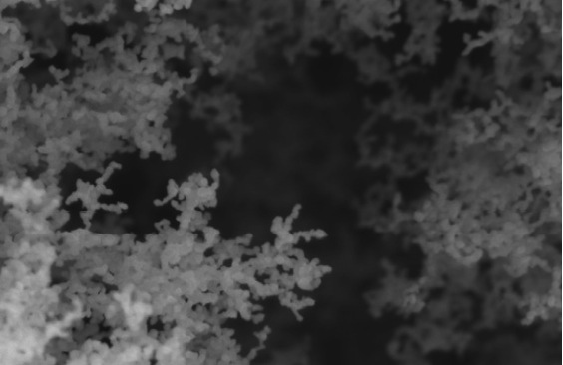}
\includegraphics[width=0.48\columnwidth]{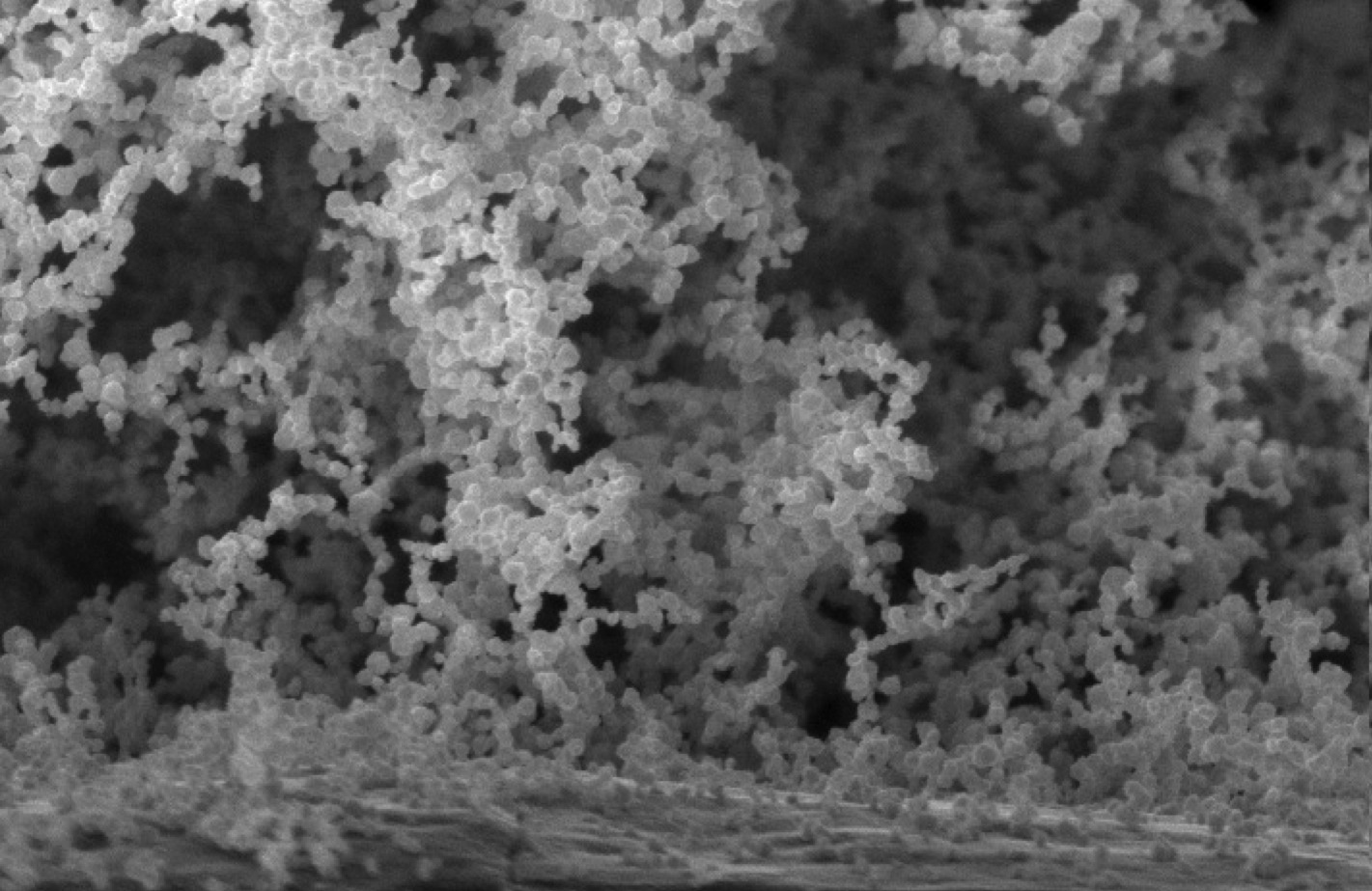}
\includegraphics[width=0.48\columnwidth]{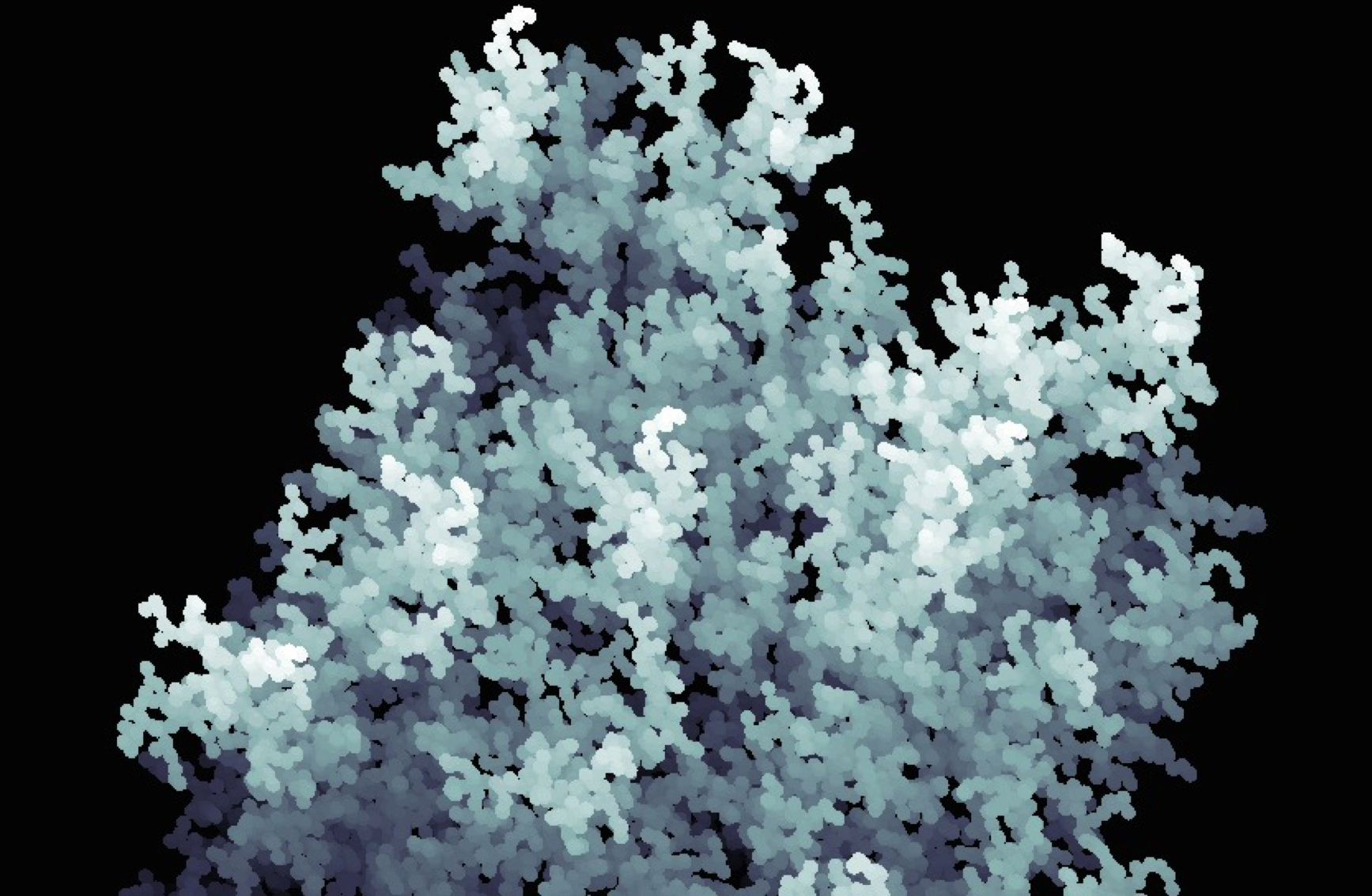}
\end{center}
\caption{Morphology of random nanostructured films.  Electron micrographs of (clockwise from lower left) stochastically nanostructured bismuth, aluminum, and copper films, and results of numerical simulation of diffusion-limited aggregation.  All SEM magnifications are the same, and each image is approximately 1.8 $\mu$m high.  
\label{SEMs}}
\end{figure}

Although much of the research into such films has used gold as the evaporant~\cite{goldblack1948,goldblack1993,goldblack2003}, we find that DLA films are easily prepared from virtually any evaporable substance. In a sense, this method achieves its generality and relative material-independence by using classical mechanics rather than chemistry to produce structures at the nanoscale.  We have used this technique to prepare DLA films of copper, silver, aluminum, iron, gold, indium, nickel, bismuth, tin, germanium, and silicon monoxide.  Films prepared using copper and bismuth appear to be stable over several years when stored in air at room temperature.  Elevated temperatures or very high optical intensities (at least two orders of magnitude higher than those used to obtain the results  presented below) can rapidly decrease the optical absorptivity of the films, presumably via oxidation or coarsening of the nanostructure.  At somewhat lower optical intensities of roughly 0.1-1~kW/m$^2$ bismuth films display photo-induced aging over the course of weeks, during which the absorptivity and photoacoustic efficiency are gradually reduced, though not to zero.  We expect that this behavior is strongly material-dependent.

Fig.~\ref{SEMs} compares electron micrographs of samples produced with this technique to numerically-generated diffusion-limited aggregates.  The qualitative agreement between the observed structure and the predicted morphology of diffusion-limited aggregates is quite good for materials like aluminum, copper, and bismuth.  
 The most striking visual aspect of films prepared in this way is their extreme blackness: as shown in Fig.~\ref{blackness}, they typically absorb 99\% of incident light across a broad range of wavelengths~\cite{Harris1967}. If the light is amplitude-modulated, the films also exhibit a strong photoacoustic response.  Although the photoacoustic phenomena modeled and presented here all occur in air, fluid-coupled ultrasound is a possible future direction of study.  For this reason we note that DLA films of copper and bismuth, though they are somewhat fragile and (in the case of copper) extremely hydrophobic, can easily be immersed in fluids without damage.  In the next section, we discuss a  simple model of the air-coupled photoacoustic properties of a nanostructured granular material.

\section{Photoacoustic Properties: Theory}
Our qualitative model of the air-coupled photoacoustic response of DLA films assumes that the films absorb nearly all light incident upon them, converting the energy into heat.  We consider the effect of this absorption on a single grain of material at the smallest (material-dependent) length scale which characterizes the structure.  Because of the diffuse fractal nature of the material, individual grains are very weakly thermally coupled to the rest of the film.  This means that the grains, which have a small heat capacity due to their small size, can change their temperature substantially and rapidly in response to sudden illumination.  When the illumination stops, our model assumes that the grains lose heat by conduction into the surrounding gas.  This is in contrast to a three-dimensional bulk metal, in which conductive losses into the rest of the material would generally be more important.  Amplitude-modulated light can thus excite a large-amplitude oscillation in the temperature of the gas near individual grains at the surface, which leads to a pressure oscillation and the emission of sound waves into the gas.   This process should occur for essentially any material, and in the 1 KHz range we have observed audible responses from other black materials such as carbon soot.  However, the efficiency of photoacoustic generation by this mechanism will fall off rapidly above some maximum frequency which depends on thermal time constants of the specific nanostructure.  For DLA films, both this maximum frequency and the expected amplitude of temperature changes are large, leading to a strong photoacoustic response in air in the ultrasound regime.  

\begin{figure}[t!]
\begin{center}
\includegraphics[width=0.9\columnwidth]{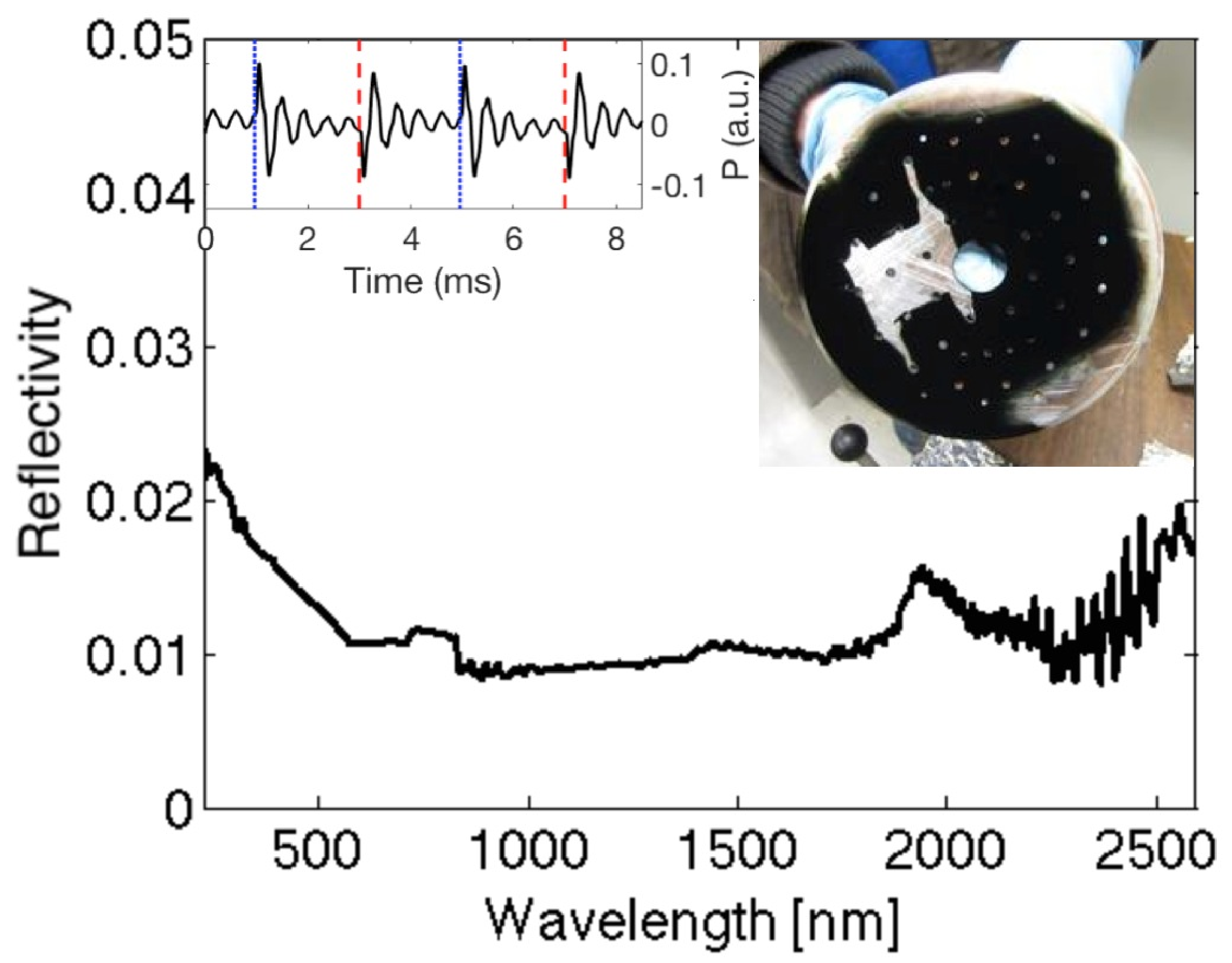}
\caption{Reflectivity versus wavelength of nanostructured copper film.  Right inset: Photograph of the measured film on a sample holder.  The bare area was covered by a sample during deposition.
Left inset: Sound pressure versus time for a film exposed to a 250 Hz square-wave optical amplitude modulation.  When the light turns on (off) at dotted (dashed) lines, the sound pressure rises (drops) rapidly. The subsequent oscillations are a consequence of ringing in the  microphone.
\label{blackness}}
\end{center}
\end{figure}

To quantitatively elucidate this simple model, we consider how quickly an individual grain in a DLA film can thermally equilibrate with its surroundings, assuming that the dominant thermal relaxation mechanism is conductivity into the neighboring gas.  The rate of thermal energy loss from a grain of metal due to conduction into the neighboring gas is
\begin{equation}
\frac{dE}{dt}= C \frac{dT}{dt} =\alpha\ \Gamma\ k_\mathrm{B}\Delta T,
\label{tempequation}
\end{equation}
where $C$ is the heat capacity of the grain, $\alpha\simeq0.15$ is a numerical factor parametrizing the degree of thermalization of a gas atom after a single collision with a surface, $\Gamma$ is the collision rate of gas atoms with the surface, and $\Delta T$ is the temperature difference between the grain and the gas.  The collision rate $\Gamma=nva$  is the number density of the gas $n$ times the RMS velocity $v$ of a gas atom times the area $a$ of the grain, which for a 10-nm-radius grain at STP is about 4 THz.  The heat capacity $C$ depends on the substance and grain size; for a 10-nm bismuth grain it is around 5$\times 10^{-18}$J/$^\circ$K.
There is therefore a material-dependent upper bound on the frequency at which such grains can change their temperature, given by $\frac{1}{\tau} = \frac{\alpha \Gamma k_B}{C}$. This result suggests that the bismuth grain can change its temperature at rates greater than 1 MHz.  

This  model suggests directions for further optimization.  A 1nm grain would respond 10 times faster than a 10nm grain (100$\times$ less area, 1000$\times$ less heat capacity).  As shown in Fig.~\ref{SEMs}, the grain size is material-dependent; aluminum, for example, produces much smaller grains than bismuth under similar deposition conditions.  Operating in helium instead of air would decrease $\tau$ by a factor of 3 due to the higher mean velocity, and operating at higher pressure would decrease $\tau$ linearly.

At frequencies below $1/\tau$, the amplitude of the photoacoustic response depends upon the amplitude of the light-induced temperature oscillation.  For light with intensity $I$ and a grain with surface area $A_{\text{grain}}$, the rate of change of temperature is 
\begin{align}
\frac{d\Delta T}{dt} = I\cdot A_{\text{grain}} - \frac{\alpha \Gamma k_B}{C}\Delta T.
\end{align}
In the steady state the optical heating rate balances the cooling rate.  For square-wave light modulation at frequencies far below $1/\tau$, we observe that the films respond with a sharp positive and negative pressure pulse at the light switching times, as shown in the inset of Fig.~\ref{blackness}. This is consistent both with our simple model and thorough theoretical analysis \cite{pressurepulses}. Possible resonant effects~\cite{PAresonancespec} are an intriguing area for future study.

This simple model ignores effects which are certainly present at some level, such as radiative loss or conductive loss into the material~\cite{VARGAS198843}.  The model does not predict the fraction of optical energy converted to acoustic energy, which we experimentally estimate to be on the order of $10^{-8}$.   For reference, this is about 3.5 orders of magnitude greater than the photoacoustic efficiency reported for a metal-coated silicon wafer in air~\cite{PAinair}. Higher efficiencies are attainable in water, although our efficiency is comparable to that reported for metallic films~\cite{Biagi2001} and nanotube-based absorbers~\cite{Baac2012} in water.  Investigation of the relative frequency-dependent efficiencies of different photoacoustic techniques is a promising direction for further research.  
\begin{figure}[t!]
\centering
\includegraphics[width=\columnwidth]{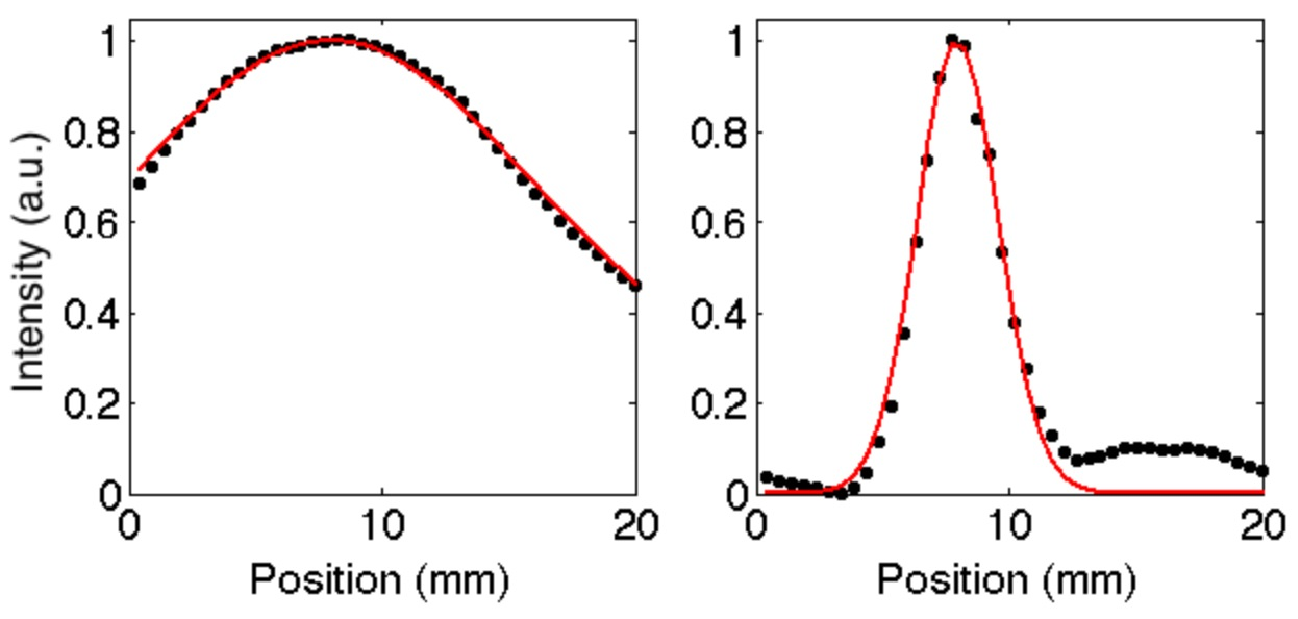}
\caption{ Diffraction-limited focused ultrasound from a DLA film. Nanostructured bismuth-black approximately 3 $\mu$m thick was deposited on a plano-concave lens with a 150-mm radius of curvature and back-illuminated with amplitude-modulated white LEDs at an optical intensity of around 140 W/m$^2$.   Acoustic intensity was measured with narrowband ultrasonic transducers: TCT40-16R for 40 kHz and SensComp 200KHF18 for 200 kHz, both using a PAR 113 preamplifier.  Points show measured acoustic intensity versus position.  Lines are gaussian curves with a frequency-dependent diffraction-limited width $w_0(\nu)$, scaled only vertically and centered on the observed spot.  Left panel is a 40KHz focus, and right panel is a 200 KHz focus.  These data were obtained by summing  2D acoustic intensity scans in one direction.}
\label{focusedsound}
\end{figure}

\section{Photoacoustic Properties: Experiment}

The photoacoustic response of all DLA films we have produced is strong enough that illumination of a flat film with an ordinary LED flashlight which uses amplitude modulation in the acoustic range produces a response audible to the human ear. More quantitatively, the strength of the photoacoustic response at 40 KHz is more than an order of magnitude stronger than that of a black graphite surface. 

The simplicity of DLA-based photoacoustic sources enables the construction of flexible and powerful acoustic devices.  The simplest incarnation of a directional DLA-based photoacoustic source is a plano-concave spherical lens with a nanostructured film deposited on the concave side. In such a device, the acoustic wavefronts generated by the DLA film inherit the  curvature of the lens surface, leading to a converging acoustic beam \cite{Baac2012}. As a proof of concept, we deposited nanostructured bismuth-black on a plano-concave lens and back-illuminated it with amplitude-modulated light to create a focused acoustic source.  Results for modulation frequencies of 40 KHz and 200 KHz (chosen based on available transducers) are presented in Fig.~\ref{focusedsound}, and compared to calculated diffraction-limited intensity curves with $1/e^2$ radius $w_0 = 2 c F / \pi \nu D$, where $c$ is the speed of sound, $F$ is the focal length, $\nu$ is the frequency, and $D$ is the lens diameter.  The foci are found to be diffraction-limited; as expected, the 200 KHz spot is 5 times smaller than the 40KHz one.

\begin{figure}[t!]
\begin{center}
\includegraphics[width=\columnwidth]{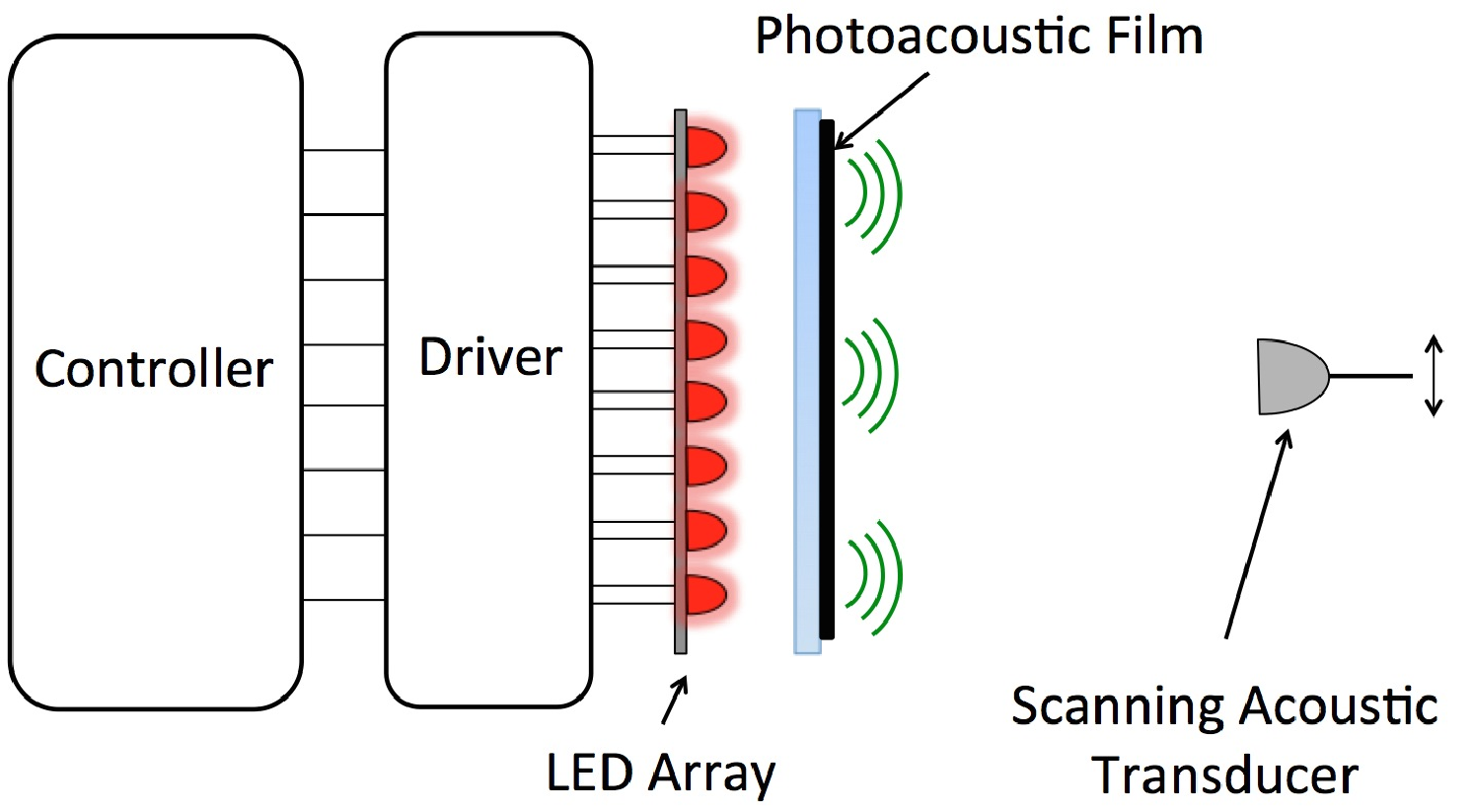}
\end{center}
\vspace{-.2in}	
\caption{Diagram of photoacoustic phased array apparatus. The field-programmable gate array  controller creates 64 amplitude-modulated voltage signals with varying phase.  These are converted to current signals by a FET driver and sent to the 8-by-8 LED array (100 mm on a side) for conversion to intensity-modulated light.  The photoacoustic film converts the amplitude-modulated light into sound  with a controllable position-dependent acoustic phase.  Propagation of the resulting wavefront creates an intensity pattern which depends on the chosen phase pattern.  After some propagation distance, the acoustic intensity as a function of position is detected by a scanning acoustic transducer.
} 
\label{PAPAdiagram}
\end{figure}

To further explore the flexibility of acoustic sources based on DLA films, we constructed an optically-addressed acoustic phased array, diagrammed in Fig.~\ref{PAPAdiagram}. A field-programmable gate array in combination with an array of FET-based current switches is used to individually control the phase of amplitude modulation for each high-power LED in an 8-by-8 square array with a pitch of 12mm.   The array is mounted directly behind a glass panel covered with nanostructured bismuth. This configuration allows for virtually arbitrary optical control of the acoustic wavefronts generated at the DLA film. Like the recently-proposed metascreen-based passive phased array~\cite{acousticphasedarray}, this device avoids the complexity and cost of a multiple-acoustic-source phased array, but in an entirely different manner.  Here the technical advantages arise from the comparative ease and flexibility of generating amplitude modulated light, as compared to direct generation of ultrasound.

\begin{figure}[t!]
\begin{center}
\includegraphics[width=\columnwidth]{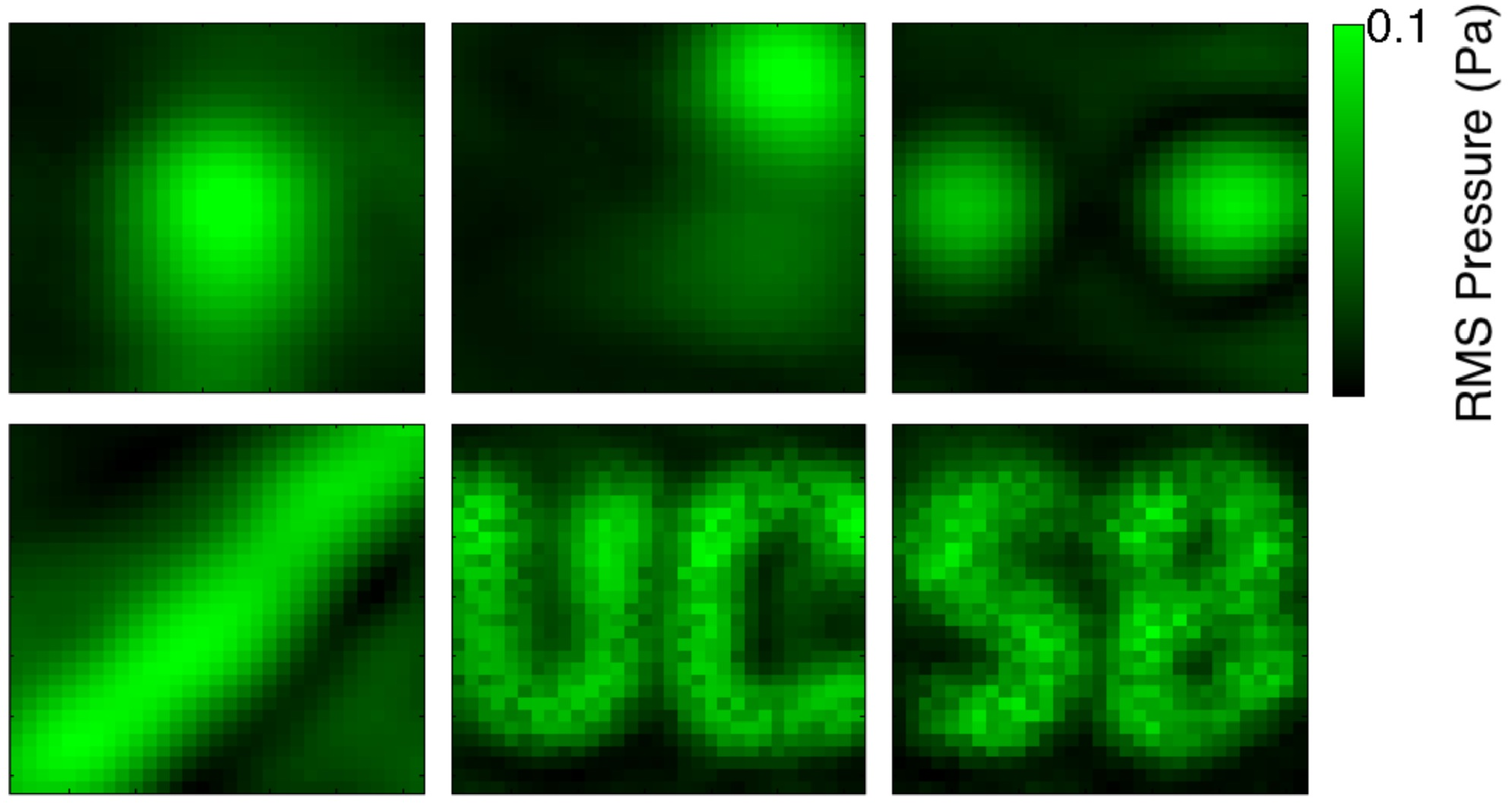}
\end{center}
\caption{Patterned acoustic pressure fields created by the optically addressed phased array, as measured by a scanning piezo transducer.  Images demonstrate the ability to create moving foci, multiple foci, and virtually arbitrary intensity patterns.  To create these images, 140 W/m$^2$ of white light was used to back-illuminate bismuth-black films with a thickness of about 3 $\mu$m.  The acoustic frequency is 40 KHz, and all images are 30 mm across except the ``UCSB'' images, which are 45mm across.  All images were taken at a distance of 70mm from the DLA film. Colorbar indicates the measured pressure for the upper-right figure.  This pressure (0.1 Pa RMS)  is representative of the typical range, though colormaps on the other subplots vary for visibility.} \label{focusing}
\end{figure}

Figure~\ref{focusing} shows a few intensity patterns created by the photoacoustic phased array and measured with a scanning acoustic transducer.  With an appropriately chosen phase configuration, the device can mimic the photoacoustic lens, creating a single focused acoustic intensity maximum as shown in the upper-left panel of the figure.  By adding a phase offset which depends linearly on position, this virtual lens can be dynamically tilted, allowing for full spatiotemporal control of the acoustic intensity maximum.  Such a translated focus is shown in the upper-middle panel of Fig.~\ref{focusing}.  Slightly more complex phase configurations can create multiple foci (upper-right panel of Fig.~\ref{focusing}) or lines of maximum acoustic intensity (lower-left panel of Fig.~\ref{focusing}).  Arbitrary spatial variation of the phase pattern can be used to create more complex waveforms, with a precision limited by acoustic diffraction and the spatial period of the LED array.  The fact that the 12mm pitch is larger than half a wavelength of 40 KHz ultrasound also limits the photoacoustic efficiency due to radiation of power into side lobes.  Time-varying phase configurations allow for ``painting'' of time-averaged acoustic intensity patterns, for example by scanning a focused spot.  The ``UCSB'' images in Fig.~\ref{focusing} were generated in this way, at a repetition rate greater than 200 Hz.

\section{Conclusion}
We have discussed, modeled, and demonstrated air-coupled photoacoustic sources based on nanostructured diffusion-limited aggregates.  These sources are inexpensive and simple to fabricate and are capable of broadband ultrasound generation. As an illustration of the device possibilities, we designed and constructed a DLA-based optically-addressed ultrasonic phased array and used it to create virtually arbitrary acoustic intensity patterns.

\begin{acknowledgments}
The authors thank Thomas Witten for a useful and interesting conversation; Carl Felten, Daeseong Kim, and Andrew Williams for experimental assistance in the construction of the acoustic phased array driver; and the Hellman Family Faculty Fellowship for support.
\end{acknowledgments}

\bibliographystyle{aip2}


\end{document}